\RequirePackage{fixltx2e}
\documentclass[11pt,a4paper]{article}
\pdfoutput=1
\usepackage{jcappub}
\usepackage[utf8]{inputenc}
\usepackage{lmodern}
\usepackage[T1]{fontenc}

\usepackage{amsfonts}
\usepackage{mathtools}
\DeclareMathAlphabet{\mathsfi}{OT1}{cmss}{m}{sl}
\DeclareMathAlphabet{\mathbfi}{OML}{cmm}{b}{it}
\usepackage{bbm}
\usepackage{mathrsfs}

\usepackage{cleveref}
\crefformat{equation}{(#2#1#3)}
\crefmultiformat{equation}{(#2#1#3}{, #2#1#3)}{, #2#1#3}{, #2#1#3)}
\crefrangeformat{equation}{(#3#1#4--#5\crefstripprefix{#1}{#2}#6)}
\renewcommand{\eqref}{\cref}

\newcommand{\total}{\mathop{}\!\mathrm{d}}
\newcommand{\eqend}[1]{\,\mathrm{#1}}
\newcommand{\mathi}{\mathrm{i}}
\newcommand{\mathe}{\mathrm{e}}
\renewcommand{\vec}[1]{\mathbfi{#1}}

\newcommand{\bra}[1]{{\left\langle{#1}\right\vert}}
\newcommand{\ket}[1]{{\left\vert{#1}\right\rangle}}

\newcommand{\abs}[1]{{\left\vert{#1}\right\vert}}
\newcommand{\laplace}{\mathop{}\!\bigtriangleup}

\newcommand{\hankel}[1]{\mathop{}\!\mathrm{H}^{(#1)}}

\newcommand{\bigo}[1]{\mathcal{O}\left({#1}\right)}

\newcommand{\lie}{\mathscr{L}}

\newcommand{\Ein}{\mathop{\mathrm{Ein}}}

\newenvironment{equations}[1][]{\subequations\ifx\relax#1\relax\else\label{#1}\fi\align\ignorespaces}{\endalign\ignorespacesafterend\endsubequations}
\makeatletter
\def\@spliteq#1{\begin{equation}\begin{split}#1\end{split}\end{equation}}
\def\spliteq{\collect@body\@spliteq}

\makeatother

\let\originalleft\left
\let\originalright\right
\renewcommand{\left}{\mathopen{}\mathclose\bgroup\originalleft}
\renewcommand{\right}{\aftergroup\egroup\originalright}

\usepackage{xspace}

\newcommand{\ie}{\textit{i.\,e.}\xspace}
\newcommand{\eg}{\textit{e.\,g.}\xspace}

\begin{document}

\title{The Weyl tensor correlator in cosmological spacetimes}

\author{Markus B. Fr{\"o}b}

\affiliation{Departament de F{\'\i}sica Fonamental, Institut de Ci{\`e}ncies del Cosmos (ICC),\\ Universitat de Barcelona (UB), C/ Mart{\'\i} i Franqu{\`e}s 1, 08028 Barcelona, Spain}
\affiliation{Institut f\"ur Theoretische Physik, Universit{\"a}t Leipzig,\\ Br{\"u}derstra{\ss}e 16, 04103 Leipzig, Germany}

\emailAdd{mfroeb@itp.uni-leipzig.de}

\abstract{We give a general expression for the Weyl tensor two-point function in a general Friedmann-Lema{\^\i}tre-Robertson-Walker spacetime. We work in reduced phase space for the perturbations, \ie, quantize only the dynamical degrees of freedom without adding any gauge-fixing term. The general formula is illustrated by a calculation in slow-roll single-field inflation to first order in the slow-roll parameters $\epsilon$ and $\delta$, and the result is shown to have the correct de Sitter limit as $\epsilon, \delta \to 0$. Furthermore, it is seen that the Weyl tensor correlation function in slow-roll does not suffer from infrared divergences, unlike the two-point functions of the metric and scalar field perturbations. Lastly, we show how to recover the usual tensor power spectrum from the Weyl tensor correlation function.}



\maketitle

\section{Introduction}

Inflation as a theory of the very early universe has been highly successful, and is compatible with all observations so far. Cosmic structures are believed to have been seeded by primordial quantum fluctuations, and information about these fluctuations can be extracted from observations of the Cosmic Microwave Background~\cite{planckxv,
planckxxiv}. The recent detection of a primordial gravitational wave signal~\cite{bicep2}, if confirmed, would provide even more support and help to uncover more details about the exact model of inflation.

It has been long known, however, that correlation functions of massless quantum fields (such as scalars or gravitational waves) in inflationary spacetimes suffer from infrared (IR) problems~\cite{fordparker1977a,fordparker1977b}. In fact, if the observed (nearly) scale-invariant spectrum (for which the equal-time two-point function is $\sim \abs{\vec{k}}^{-3}$) persists for arbitrary small momenta $\vec{k}$, the Fourier transform of the two-point function, which in the IR then reads $\sim \int \abs{\vec{k}}^{-3} \total^3 k$, diverges logarithmically. Many approaches have been suggested to deal with this divergence~\cite{riottosloth2008,burgessleblondholmanshandera2009,rajaramankumarleblond2010,garbrechtrigopoulos2011,giddingssloth2011b,gerstenlauerhebeckertasinato2011} (see also the review article~\cite{seery2010}), among the simplest an IR cutoff $\xi$ placed at small momentum. One could try to make sense of such an cutoff, \eg, by taking it to be the inverse of the size of our inflationary patch at the beginning of inflation, but the result of calculations may depend strongly on the exact choice. A related issue is the appearance of secular terms for massless fields in the natural (Bunch-Davies) vacuum. These terms usually grow like logarithms of the scale factor, and could thus dominate the late-time behaviour and spoil perturbation theory. One would think that proper physical observables (which are intrinsically local or at least of finite extent, since measurements are finite) should not depend on such arbitrary cutoffs, and that secular terms should also only be regarded as physical if they appear in such observables.

Many calculations have been done in de Sitter spacetime (where the relevant part for inflation is the Poincar{\'e} patch), and which can be seen as an inflationary spacetime where inflation is just driven by a cosmological constant $\Lambda$. There it has been shown that the IR divergences of the graviton two-point function are pure gauge, in the sense that they can be set to zero in any arbitrarily large (but finite) region by a linearised gauge transformation, and the explicit form of this transformation has been derived~\cite{higuchimarolfmorrison2011}. This is in concordance with the fact that correlation functions of gauge-invariant observables, such as the linearised Weyl tensor, do not show any IR divergence nor secular growth when an IR cutoff is introduced. The Weyl tensor correlation function in de Sitter space has been derived many times~\cite{kouris2001,kouris2001corr,morawoodard2012,moratsamiswoodard2012a,froebrouraverdaguer2014}, and agrees for the Bunch-Davies vacuum state in the Poincar{\'e} patch (where IR divergences are present for the graviton propagator) and the Euclidean vacuum state in global de Sitter space (where no IR divergences arise). One may thus conclude that --- at least at tree level --- these IR divergences are unphysical. The situation is less clear when interactions are taken into account; it has been shown that the inclusion of loops of conformal matter~\cite{froebrouraverdaguer2012,froebrouraverdaguer2014} does not change the above picture, \ie, local and gauge-invariant observables do not show any IR divergence nor secular growth, and extensions to other matter fields are currently being investigated~\cite{frv2015}. It thus seems that the only effect of the graviton IR divergence could possibly arise from internal graviton lines or graviton self-interactions. While early calculations~\cite{tsamiswoodard1996,tsamiswoodard1997,kahyaonemliwoodard2010} seem to find an effect (manifesting itself in the appearance of secular logarithms), the correctness of the results has been challenged~\cite{garrigatanaka2008,pimentelsenatorezaldarriaga2012,senatorezaldarriaga2013,assassibaumanngreen2013}, and the issue is far from settled. Part of the difficulty of quantifying such effects stems from the fact that one needs to consider local and gauge-invariant observables --- objects that could actually be measured ---, and the construction of such observables is notoriously difficult.

Without diminishing the merit of the previously mentioned calculations, the geometry of the primordial universe is however only approximately described by de Sitter space, since the primordial expansion is not exactly exponential. A better approximation is given by considering a general Friedmann-Lema{\^\i}tre-Robertson-Walker universe, with the scale factor describing an almost exponential expansion. Such backgrounds are generated, \eg, in so-called slow-roll models where inflation is driven by one (or more) scalar fields slowly rolling down a potential. In these cosmological spacetimes, infrared divergences generally become worse~\cite{janssenprokopec2008,giddingssloth2011a,xuegaobrandenberger2012}, and it is thus important to try to carry over the insights obtained in de Sitter space. It has been argued~\cite{urakawatanaka2010,tanakaurakawa2013,tanakaurakawa2014} that also in cosmological spacetimes the IR divergences disappear in the same way as in de Sitter space, namely when one considers local and gauge-invariant observables. The present article is a step in this direction: quantizing metric and scalar field perturbations, we derive the Weyl tensor correlation function for these perturbations for an arbitrary flat FLRW background at tree level. For definiteness, we consider a single-field inflationary model with the standard Einstein-Hilbert action for gravity, but our considerations could be easily extended to multi-field models and/or models of modified gravity as well.

The article is structured as follows: In the first section~\ref{sec_metricdecomp}, we consider metric and scalar field perturbations around a flat FLRW background, and separate them into gauge-invariant and gauge-dependent parts. In the next section~\ref{sec_action}, we expand the action to second order in perturbations, where only the gauge-invariant parts of the perturbations remain. Afterwards, we separate dynamical and constrained degrees of freedom and recover the well-known dynamical degrees of freedom: the tensor part of the graviton and the Mukhanov-Sasaki variable. The reason for rederiving this well-known result lies in the method used, which is a) easily generalised to other models including models where vector or scalar parts of the graviton are dynamical, and b) permits to identify gauge-invariant observables in a very simple manner: exactly those observables (or correlation functions) which only depend on the gauge-invariant parts of the perturbation are gauge-invariant (locality is usually evident). In section~\ref{sec_weyl}, the linearised Weyl tensor is expressed in terms of the metric perturbations and seen to be gauge-invariant. The dynamical graviton and scalar perturbations are then quantized, and a general formula expressing the Weyl correlation function in terms of the mode functions of the perturbations is given, and a simpler formula for the four-dimensional Wightman function is derived. In the last section~\ref{sec_slowroll}, we apply said formula to the case of slow-roll inflation and calculate the Weyl tensor correlation function to first order in the slow-roll parameters. It is seen that the IR divergences, which appear for the graviton and scalar two-point functions, do not contribute to the Weyl tensor correlator, confirming thus the conjectured resolution of the IR problem in this case. The de Sitter limit of the result is taken and shown to coincide with the previous results~\cite{kouris2001,kouris2001corr,morawoodard2012,moratsamiswoodard2012a,froebrouraverdaguer2014}. Lastly, in section~\ref{sec_powerspectrum} we show how to recover more familiar observables such as the power spectrum from the Weyl tensor correlator, and clarify how a nearly scale-invariant spectrum arises, even if the Weyl tensor is IR-safe. Our conventions are the ``+++'' convention of~\cite{mtw}, Greek indices are spacetime ones and Latin indices indicate purely spatial components.

\section{Decomposition of metric perturbations}
\label{sec_metricdecomp}

Let us consider the metric perturbations around a flat FLRW background. We decompose the full metric as
\begin{equation}
\tilde{g}_{\mu\nu} = \tilde{g}^{(0)}_{\mu\nu} + \tilde{g}^{(1)}_{\mu\nu} \eqend{,}
\end{equation}
where the background metric $\tilde{g}^{(0)}_{\mu\nu}$ is given by
\begin{equation}
\label{flrw}
\tilde{g}^{(0)}_{\mu\nu} = a^2(\eta) \eta_{\mu\nu}
\end{equation}
where the scale factor $a$ depends on conformal time $\eta$, derivatives with respect to conformal time are denoted by a prime, and $\eta_{\mu\nu}$ is the flat Minkowski metric. For the derivatives of the scale factor we introduce the Hubble parameter $H$ via
\begin{equation}
\label{hubble_def}
H \equiv \frac{a'}{a^2} \eqend{.}
\end{equation}
The perturbation $\tilde{g}^{(1)}_{\mu\nu}$ is also rescaled to give
\begin{equation}
h_{\mu\nu} \equiv a^{-2} \tilde{g}^{(1)}_{\mu\nu} \eqend{,}
\end{equation}
so that we obtain the usual FLRW form $\tilde{g}_{\mu\nu} = a^2 ( \eta_{\mu\nu} + h_{\mu\nu} )$. Singling out the time coordinate, the metric perturbation can be decomposed into its irreducible parts with respect to spatial transformations, and we obtain
\begin{equations}[irreducible]
h_{00} &= s_1 \eqend{,} \\
h_{0k} &= v^\text{T1}_k + \partial_k s_2 \eqend{,} \\
h_{kl} &= h^\text{TT}_{kl} + 2 \partial_{(k} v^\text{T2}_{l)} + \left( \partial_k \partial_l - \frac{\delta_{kl} \laplace}{n-1} \right) s_3 + \delta_{kl} s_4 \eqend{.}
\end{equations}
Note that we consider an $n$-dimensional background, having in mind the use of dimensional regularization for an eventual later calculation of loop corrections. The tensor perturbation $h^\text{TT}_{kl}$ is transverse and traceless, $h^\text{TT}_{kk} = \partial_k h^\text{TT}_{kl} = 0$, while the vectors are transverse $\partial_k v^\text{T1}_k = \partial_k v^\text{T2}_k = 0$ (since the spatial metric is the identity, we do not make a distinction between upper and lower spatial indices, which are summed over regardless of their position). Under an infinitesimal coordinate transformation with parameter $\tilde{\xi}^\mu$, we have
\begin{equation}
\delta h_{\mu\nu} = a^{-2} \lie_{\tilde{\xi}} \left( a^2 \eta_{\mu\nu} \right) = 2 \partial_{(\mu} \xi_{\nu)} - 2 H a \eta_{\mu\nu} \xi_0 \eqend{,}
\end{equation}
where we rescaled $\tilde{\xi}^\mu$ as
\begin{equation}
\tilde{\xi}_\mu = a^2 \xi_\mu \eqend{.}
\end{equation}
Inserting the decomposition~\eqref{irreducible}, we obtain the transformation rules for the irreducible components
\begin{equations}
\delta s_1 &= 2 \xi_0' + 2 H a \xi_0 \eqend{,} \\
\delta s_2 &= \xi_0 + \frac{\partial_k}{\laplace} \xi_k' \eqend{,} \\
\delta s_3 &= 2 \frac{\partial_k}{\laplace} \xi_k \eqend{,} \\
\delta s_4 &= \frac{2}{(n-1)} \partial_k \xi_k - 2 H a \xi_0 \eqend{,} \\
\delta v^\text{T1}_k &= \xi_k' - \frac{\partial_k \partial_l}{\laplace} \xi_l' \eqend{,} \\
\delta v^\text{T2}_k &= \xi_k - \frac{\partial_k \partial_l}{\laplace} \xi_l \eqend{,} \\
\delta h^\text{TT}_{kl} &= 0 \eqend{.}
\end{equations}
Note that in this section we only consider perturbations and gauge transformations of finite extent, such that the inverse of the Laplace operator (with vanishing boundary conditions) is well defined. The tensor is gauge-invariant by itself, and we define
\begin{equation}
\label{h1_def}
H_{kl} \equiv h^\text{TT}_{kl} \eqend{.}
\end{equation}
From the two vectors, we can form one gauge-invariant combination
\begin{equation}
\label{v1_def}
V_k \equiv v^\text{T1}_k - v^{\text{T2}\prime}_k \eqend{,}
\end{equation}
and there are two gauge-invariant scalars,
\begin{equations}
S \equiv s_1 - ( 2 s_2 - s_3' )' - H a ( 2 s_2 - s_3' ) \eqend{,} \\
\Sigma \equiv s_4 - \frac{1}{n-1} \laplace s_3 + H a ( 2 s_2 - s_3' ) \eqend{.}
\end{equations}
Furthermore, the remaining components can be organizied into a vector $\tilde{X}^\mu$
\begin{equations}
\tilde{X}^0 &= \frac{1}{2} s_3' - s_2 \eqend{,} \\
\tilde{X}^k &= \eta^{kl} \left( v^\text{T2}_l + \frac{1}{2} \partial_l s_3 \right) \eqend{,}
\end{equations}
which has the simple gauge transformation
\begin{equation}
\delta \tilde{X}^\mu = \tilde{\xi}^\mu \eqend{.}
\end{equation}
Rescaling $\tilde{X}_\mu = a^2 X_\mu$, the covariant components read
\begin{equations}
X_0 &= - \frac{1}{2} s_3' + s_2 \eqend{,} \\
X_k &= v^\text{T2}_k + \frac{1}{2} \partial_k s_3 \eqend{,}
\end{equations}
and transform as $\delta X_\mu = \xi_\mu$. We thus have
\begin{equations}[h_decomp]
h_{00} &= S + 2 X_0' + 2 H a X_0 \eqend{,} \\
h_{0k} &= V_k + X_k' + \partial_k X_0 \eqend{,} \\
h_{kl} &= H_{kl} + \delta_{kl} \Sigma + 2 \partial_{(k} X_{l)} - 2 H a \delta_{kl} X_0 \eqend{.}
\end{equations}

Apart from the metric peturbations, we can also introduce matter fields. For a scalar $\tilde{\phi}$, which we decompose into a background value $\phi$ and a perturbation $\psi$, we obtain the change
\begin{equation}
\label{psi_change}
\delta \psi = \lie_{\tilde{\xi}} \phi = \tilde{\xi}^\mu \partial_\mu \phi
\end{equation}
under an infinitesimal coordinate transformation. The FLRW background~\eqref{flrw} can be generated by the stress tensor of such a scalar field, then called the inflaton, which only depends on time. In this case, the change of its perturbation~\eqref{psi_change} is given by $\delta \psi = - \xi_0 \phi'$, and thus the combination
\begin{equation}
\label{psi_def}
\Psi \equiv \psi + X_0 \phi'
\end{equation}
is clearly gauge invariant. Similar combinations can be found for other matter fields.

\section{The gauge-invariant action}
\label{sec_action}

\subsection{Deriving the action}

For definiteness, we consider a simple single-field inflationary model, given by the standard Einstein-Hilbert action for gravity and the action of a minimally coupled scalar field
\begin{equation}
\label{action}
I = \frac{1}{\kappa^2} \int \tilde{R} \sqrt{-\tilde{g}} \total^n x - \frac{1}{2} \int \left[ \tilde{g}^{\mu\nu} ( \partial_\mu \tilde{\phi} ) ( \partial_\nu \tilde{\phi} ) + V(\tilde{\phi}) \right] \sqrt{-\tilde{g}} \total^n x \eqend{.}
\end{equation}
Here $\kappa^2 = 16 \pi G_\text{N}$ with Newton's constant $G_\text{N}$, and $V(\tilde{\phi})$ is an arbitrary potential for the inflaton $\tilde{\phi}$. The expansion of this action in perturbations can be done using the formulas from appendix~\ref{app_expansion}, but it is less tedious to use a computer algebra system such as xAct~\cite{xact}. Integrating by parts, at linear order we obtain
\begin{spliteq}
I^{(1)} &= \frac{1}{2 \kappa^2} \int \left[ - (n-1) (n-2) H^2 a^2 + \frac{1}{2} \kappa^2 ( \phi' )^2 + \frac{1}{2} a^2 \kappa^2 V(\phi) \right] h_{00} a^{n-2} \total^n x \\
&\quad+ \frac{1}{2 \kappa^2} \int \left[ 2 (n-2) H' a + (n-1) (n-2) H^2 a^2 + \frac{1}{2} \kappa^2 ( \phi' )^2 - \frac{1}{2} a^2 \kappa^2 V(\phi) \right] h_{kk} a^{n-2} \total^n x \\
&\quad- \int \left[ \phi'' + (n-2) H a \phi' + \frac{1}{2} a^2 V'(\phi) \right] \psi a^{n-2} \total^n x \eqend{.}
\end{spliteq}
Varying the action with respect to $h_{00}$, $h_{kk}$ and $\psi$, we obtain the background equations of motion (the Friedmann equations for this case), which we write in the form
\begin{equations}[friedmann]
\label{friedmann_v} \kappa^2 V(\phi) &= 2 (n-2) \left( (n-1) H^2 + \frac{H'}{a} \right) \eqend{,} \\
\kappa^2 ( \phi' )^2 &= - 2 (n-2) H' a \eqend{,} \\
\phi'' &= - (n-2) H a \phi' - \frac{1}{2} a^2 V'(\phi) \eqend{.}
\end{equations}
Note that the last equation follows from the first two. We will use these equations in the following to replace the left-hand sides by the right-hand ones whenever possible, without stating it explicitly each time. An equation which is also needed follows by differentiating~\eqref{friedmann_v} and reads
\begin{equation}
\kappa^2 V'(\phi) \phi' = 2 (n-2) \left[ (2n-3) H' H + \frac{H''}{a} \right] \eqend{.}
\end{equation}

We now pass to quadratic order. Decomposing the metric perturbation as in equation~\eqref{h_decomp}, we obtain
\begin{spliteq}
\label{action_2}
I^{(2)} &= \frac{1}{4 \kappa^2} \int \bigg[ H_{kl}' H_{kl}' + H_{kl}' \laplace H_{kl}' - 2 V_k \laplace V_k \bigg] a^{n-2} \total^n x \\
&\quad- \frac{n-2}{4 \kappa^2} \int \left[ ( H' a + (n-1) H^2 a^2 ) S^2 + 2 (n-1) H a \Sigma' S \right] a^{n-2} \total^n x \\
&\quad- \frac{n-2}{4 \kappa^2} \int \left[ (n-1) \left( \Sigma' \right)^2 + (n-3) \Sigma \laplace \Sigma - 2 S \laplace \Sigma \right] a^{n-2} \total^n x \\
&\quad+ \frac{1}{2} \int S \left[ \phi' \Psi' + \frac{1}{2} a^2 V'(\phi) \Psi \right] a^{n-2} \total^n x \\
&\quad+ \frac{1}{2} \int \left[ \left( \Psi' \right)^2 + \Psi \laplace \Psi - \frac{1}{2} a^2 V''(\phi) \Psi^2 - \phi' \Psi (n-1) \Sigma' \right] a^{n-2} \total^n x \eqend{.}
\end{spliteq}
We see that only the gauge-invariant combinations remain and that all terms involving $X_\mu$ have cancelled, as was to be expected from the gauge invariance of the original action.

\subsection{Separating dynamical and constrained degrees of freedom}

In General Relativity, not all parts of the metric perturbation and thus not all gauge-invariant combinations are dynamical. In the ADM formalism~\cite{adm} this is very clear: lapse and shift are merely Lagrange multipliers in the action, and only the tensor part of the metric perturbations (and the inflaton) are dynamical. In our approach, one can clearly see from the quadratic part of the action~\eqref{action_2} that the vector $V_k$ and the scalar $S$ are constrained and their equation of motion is algebraic. However, it seems that both $\Psi$ and $\Sigma$ are dynamical, since for both a term quadratic in time derivatives exists in the action. The constraint fields can be eliminated by deriving their equations of motion and substituting the solutions back into the action, and we will see that after this elimination the action correctly describes the inflationary dynamical degrees of freedom. The equation of motion for the vector reads
\begin{equation}
\label{v_eom}
\laplace V_k = 0 \eqend{,}
\end{equation}
and substituting it back into the action~\eqref{action_2} the vector part simply vanishes. In contrast, for the scalar $S$ we obtain the more complicated expression
\begin{equation}
\label{s_eom}
( H' a + (n-1) H^2 a^2 ) S = - (n-1) H a \Sigma' + \laplace \Sigma + \frac{\kappa^2}{(n-2)} \left( \phi' \Psi' + \frac{1}{2} a^2 V'(\phi) \Psi \right) \eqend{.}
\end{equation}
Substituting it back into the action, we obtain
\begin{spliteq}
\label{action_2_subst1}
&\frac{1}{4 \kappa^2} \int \bigg[ H_{kl}' H_{kl}' + H_{kl} \laplace H_{kl} \bigg] a^{n-2} \total^n x + \frac{n-1}{2} \int \frac{( \Psi' )^2}{H' a + (n-1) H^2 a^2} H^2 a^n \total^n x \\
&\quad- \frac{1}{4} \int \frac{(n-1) H^2 a^4 V''(\phi) - \frac{\kappa^2}{2 (n-2)} a^4 [ V'(\phi) ]^2 + 2 H a^2 ( H'' + (2n-3) H' H a )}{H' a + (n-1) H^2 a^2} \Psi^2 a^{n-2} \total^n x \\
&\quad+ \frac{1}{2} \int \frac{- (n-1) H a \Sigma' + \laplace \Sigma}{H' a + (n-1) H^2 a^2} \left[ \phi' \Psi' + \frac{1}{2} a^2 V'(\phi) \Psi \right] a^{n-2} \total^n x \\
&\quad+ \frac{n-2}{4 \kappa^2} \int \frac{\left[ - (n-1) H' a ( \Sigma' )^2 + ( \laplace \Sigma )^2 \right]}{H' a + (n-1) H^2 a^2} a^{n-2} \total^n x + \frac{n-2}{2 \kappa^2} \int \Sigma \laplace \Sigma a^{n-2} \total^n x \\
&\quad+ \frac{1}{2} \int \Psi \laplace \Psi a^{n-2} \total^n x - \frac{n-1}{2} \int \Sigma' \phi' \Psi a^{n-2} \total^n x \eqend{.}
\end{spliteq}
It still seems that there are two dynamical scalars, but there is a cross term $\Sigma' \Psi'$ which we must remove by diagonalizing before we can make a definite statement. This can be done by defining the Sasaki-Mukhanov variable~\cite{mukhanovfeldmanbrandenberger1992}
\begin{equation}
\label{q_def}
Q \equiv \frac{2 H a}{\phi'} \Psi - \Sigma
\end{equation}
and replacing $\Psi$ by $Q$, so that we obtain
\begin{spliteq}
\label{action_2_subst2}
I^{(2)} &= \frac{1}{4 \kappa^2} \int \bigg[ H_{kl}' H_{kl}' + H_{kl} \laplace H_{kl} \bigg] a^{n-2} \total^n x - \frac{n-2}{4 \kappa^2} \int \bigg[ \left( Q' \right)^2 + Q \laplace Q \bigg] \frac{H'}{H^2} a^{n-3} \total^n x \\
&\quad+ \frac{n-2}{4 \kappa^2} \int \frac{\left( \laplace \Sigma - \frac{H'}{H} Q' \right)^2}{H' a + (n-1) H^2 a^2} a^{n-2} \total^n x \eqend{.}
\end{spliteq}
The equation of motion for $\Sigma$ is now elliptic, namely
\begin{equation}
\label{sigma_eom}
\laplace \Sigma = \frac{H'}{H} Q' \eqend{,}
\end{equation}
and thus the action contains only the well-known two degrees of freedom, a tensorial one (which in four dimensions has two polarizations) and a scalar one.

\section{The Weyl tensor}
\label{sec_weyl}

\subsection{Expression in dynamical variables}

Since the Weyl tensor is conformally invariant, it vanishes in conformally flat spacetimes, so that for its background value we have $\tilde{C}^{(0)}_{\mu\nu\rho\sigma} = 0$. Therefore its perturbation $\tilde{C}^{(1)}_{\mu\nu\rho\sigma}$ is gauge-invariant under infinitesimal coordinate transformations (to first order), since we have
\begin{equation}
\delta \tilde{C}^{(1)}_{\mu\nu\rho\sigma} = \lie_{\tilde{X}} \tilde{C}^{(0)}_{\mu\nu\rho\sigma} = 0 \eqend{.}
\end{equation}
This also entails that in its expansion to linear order in metric perturbations, using the decomposition~\eqref{h_decomp}, the vector $X_\mu$ cannot appear. Using again either the expansions from appendix~\ref{app_expansion} or a CAS, the explicit expression is given by
\begin{equation}
\label{weyl_tilde}
\tilde{C}^{(1)}{}^{\mu\nu}{}_{\rho\sigma} = a^{-2} C^{\mu\nu}{}_{\rho\sigma}
\end{equation}
with
\begin{spliteq}
C^{\mu\nu}{}_{\rho\sigma} = - 2 \bigg[ &\partial^{[\mu} \partial_{[\rho} h_{\sigma]}^{\nu]} + \frac{1}{n-2} \delta^{[\mu}_{[\rho} \left( \partial^{\nu]} \partial_\alpha h_{\sigma]}^\alpha + \partial_{\sigma]} \partial^\alpha h^{\nu]}_\alpha - \partial^2 h^{\nu]}_{\sigma]} - \partial^{\nu]} \partial_{\sigma]} h \right) \\
&\quad+ \frac{1}{(n-1) (n-2)} \delta^\mu_{[\rho} \delta_{\sigma]}^\nu \left( \partial^2 h - \partial_\alpha \partial_\beta h^{\alpha\beta} \right) \bigg] \eqend{.}
\end{spliteq}
As expected, this is a simple rescaling from the flat space expression. Inserting the decomposition~\eqref{h_decomp}, we obtain
\begin{equations}[weyl_decomp]
2 (n-2) C^{0j}{}_{0l} &= (n-3) H^j_l{}'' + \laplace H^j_l - (n-3) \left( \partial^j V_l' + \partial_l V^j{}' \right) - \frac{n-3}{n-1} \Pi^j_l \laplace \left( S + \Sigma \right) \eqend{,} \\
C^{0j}{}_{kl} &= \partial_{[k} H_{l]}^j{}' - \partial^j \partial_{[k} V_{l]} + \frac{1}{n-2} \delta^j_{[k} \laplace V_{l]} \eqend{,} \\
\begin{split}
C^{ij}{}_{kl} &= - 2 \partial^{[i} \partial_{[k} H^{j]}_{l]} + \frac{2}{n-2} \delta^{[i}_{[k} \left( \partial^2 H^{j]}_{l]} + \partial^{j]} V_{l]}' + \partial_{l]} V^{j]}{}' \right) \\
&\quad+ \frac{2}{(n-1)(n-2)} \Pi^{[i}_{[k} \delta^{j]}_{l]} \laplace \left( S + \Sigma \right) \eqend{.}
\end{split}
\end{equations}
with the traceless projection operator
\begin{equation}
\label{proj_tr_def}
\Pi_{kl} = \delta_{kl} - (n-1) \frac{\partial_k \partial_l}{\laplace} \eqend{.}
\end{equation}
To linear order, the Weyl tensor is thus a local and gauge-invariant observable. All other index combinations are either related by the Weyl tensor symmetries to the ones above, or vanish (\eg, we have $C^{ij}{}_{0l} = - \eta_{lm} \eta^{ip} \eta^{jq} C^{0m}{}_{pq}$). Inserting the equations of motion for the constrained degrees of freedom~\eqref{v_eom}, \eqref{s_eom} and \eqref{sigma_eom} and replacing $\Psi$ by $Q$~\eqref{q_def}, this reduces to
\begin{equations}[weyl_dyn]
\begin{split}
2 (n-2) C^{0j}{}_{0l} &= (n-3) H^j_l{}'' + \laplace H^j_l - \frac{n-3}{n-1} \Pi^j_l \left[ \left( \epsilon Q' \right)' - \epsilon \laplace Q \right] \eqend{,}
\end{split} \\
C^{0j}{}_{kl} &= \partial_{[k} H_{l]}^j{}' \eqend{,} \\
(n-2) C^{ij}{}_{kl} &= - 2 (n-2) \partial^{[i} \partial_{[k} H^{j]}_{l]} + 2 \delta^{[i}_{[k} \partial^2 H^{j]}_{l]} + \frac{2}{n-1} \Pi^{[i}_{[k} \delta^{j]}_{l]} \left[ \left( \epsilon Q' \right)' - \epsilon \laplace Q \right] \eqend{.}
\end{equations}
where we defined the deceleration parameter $\epsilon$ by
\begin{equation}
\label{eps_def}
\epsilon \equiv - \frac{H'}{H^2 a}
\end{equation}
to shorten the resulting expressions. Possibly contrary to expectations, the Weyl tensor also involves the Sasaki-Mukhanov variable $Q$, which is basically introduced through the scalar constraint equations~\eqref{s_eom} and \eqref{sigma_eom}. To calculate the Weyl tensor correlation function, we must thus quantize both the tensor and scalar perturbations.

\subsection{Quantization of perturbations}

The equations of motion for the dynamical perturbations which are derived from the action~\eqref{action_2_subst2} read
\begin{equations}[dyn_eom]
\label{hkl_eom} H_{kl}'' + (n-2) H a H_{kl}' - \laplace H_{kl} &= 0 \eqend{,} \\
\label{q_eom} Q'' + \left( \frac{\epsilon'}{\epsilon} + (n-2) H a \right) Q' - \laplace Q &= 0 \eqend{.}
\end{equations}
Canonical quantization proceeds by promoting the fields to operators and deriving a complete set of mode solutions to these equations in the form of
\begin{equations}[modefunctions]
H_{kl}(\eta,\vec{x}) &= \int e_{kl}(\vec{p}) a_\vec{p} f_{\vec{p}}(\eta) \mathe^{\mathi \vec{p} \vec{x}} \frac{\total^{n-1} p}{(2\pi)^{n-1}} + \text{h.c.} \eqend{,} \\
Q(\eta,\vec{x}) &= \int b_\vec{p} q_{\vec{p}}(\eta) \mathe^{\mathi \vec{p} \vec{x}} \frac{\total^{n-1} p}{(2\pi)^{n-1}} + \text{h.c.} \eqend{,}
\end{equations}
where $e_{kl}(\vec{p})$ is a polarization tensor which enforces transverse- and tracelessness
\begin{equation}
p_k e_{kl}(\vec{p}) = e_{kk}(\vec{p}) = 0 \eqend{,}
\end{equation}
and where $a^\dagger_\vec{p}$ and $a_\vec{p}$ resp. $b^\dagger_\vec{p}$ and $b_\vec{p}$ are creation and annihilation operators. The mode functions $f_{\vec{p}}(\eta)$ and $q_{\vec{p}}(\eta)$ are normalised by demanding that the standard commutation relations
\begin{equation}
[ a(\vec{p}), a^\dagger(\vec{q}) ] = (2\pi)^{n-1} \delta^{n-1}(\vec{p}-\vec{q})
\end{equation}
lead to the canonical commutation relations
\begin{equations}[ccr]
[ Q(\eta,\vec{x}), \pi(\eta,\vec{y}) ] &= \mathi \delta^{n-1}(\vec{x}-\vec{y}) \eqend{,} \\
[ H_{kl}(\eta,\vec{x}), \pi_{ij}(\eta,\vec{y}) ] &= \mathi \left( P_{k(i} P_{j)l} - \frac{1}{n-2} P_{kl} P_{ij} \right) \delta^{n-1}(\vec{x}-\vec{y})
\end{equations}
with the canonical momenta obtained from the action~\eqref{action_2_subst2}
\begin{equations}
\pi_{kl} &= \frac{\delta}{\delta H_{kl}'} I^{(2)} = \frac{a^{n-2}}{2 \kappa^2} H_{kl}' \eqend{,} \\
\pi &= \frac{\delta}{\delta Q'} I^{(2)} = \frac{n-2}{2 \kappa^2} \epsilon H^2 a^{n-2} Q'
\end{equations}
and the projection operator
\begin{equation}
\label{proj_def}
P_{kl} = \delta_{kl} - \frac{\partial_k \partial_l}{\laplace} \eqend{.}
\end{equation}
Note that the complicated tensor factor for the tensor perturbations derives from the necessity to enforce transverse- and tracelessness, which is necessary because we work in a reduced phase-space formalism where these conditions apply directly for the operators. In contrast, if one would quantize by adding a gauge-fixing term to the original action~\eqref{action}, this factor would reduce to just $\delta_{k(i} \delta_{j)l}$. Up to phase factors, this gives the normalisation
\begin{equations}[normalisation]
f_{\vec{p}}(\eta) f^{*\prime}_{\vec{p}}(\eta) - f'_{-\vec{p}}(\eta) f^*_{-\vec{p}}(\eta) &= \mathi \kappa^2 a^{2-n} \eqend{,} \\
e_{kl}(\vec{p}) e_{ij}(\vec{p}) &= 2 P_{k(i} P_{j)l} - \frac{2}{n-2} P_{kl} P_{ij} \eqend{,} \\
q_{\vec{p}}(\eta) q^{*\prime}_{\vec{p}}(\eta) - q'_{-\vec{p}}(\eta) q^*_{-\vec{p}}(\eta) &= \mathi \frac{2 \kappa^2}{(n-2)} \frac{a^{2-n}}{\epsilon} \eqend{.}
\end{equations}

A quantum state $\ket{0}$ is then defined by requiring that the operators $a_\vec{p}$ and $b_\vec{p}$ annihilate this state, $a_\vec{p} \ket{0} = b_\vec{p} \ket{0} = 0$ for all $\vec{p}$. It is well known that this state and its interpretation depends on the choice of mode functions: with respect to the particles created by $a^\dagger_\vec{p}$ and $b^\dagger_\vec{p}$ this state is a vacuum state as implied by the notion $\ket{0}$, but by no means it must necessarily reduce to the standard Minkowski vacuum in the appropriate limit. Only if the mode functions reduce to positive-frequency exponentials in this limit do we recover the standard Minkowski vacuum state. The correlation functions in this quantum state are then given by
\begin{equation}
\label{hkl_corr}
\bra{0} H_{ij}(\eta, \vec{x}) H_{kl}(\eta', \vec{x}') \ket{0} = \int \left( 2 P_{k(i} P_{j)l} - \frac{2}{n-2} P_{kl} P_{ij} \right) f_{\vec{p}}(\eta) f^*_{\vec{p}}(\eta') \mathe^{\mathi \vec{p} (\vec{x}-\vec{x}')} \frac{\total^{n-1} p}{(2\pi)^{n-1}}
\end{equation}
for the tensor perturbations and
\begin{equation}
\label{q_corr}
\bra{0} Q(\eta, \vec{x}) Q(\eta', \vec{x}') \ket{0} = \int q_{\vec{p}}(\eta) q^*_{\vec{p}}(\eta') \mathe^{\mathi \vec{p} (\vec{x}-\vec{x}')} \frac{\total^{n-1} p}{(2\pi)^{n-1}}
\end{equation}
for the Sasaki-Mukhanov variable. At tree level, there are no cross correlations. As usual, time-ordered expectation values are defined by
\begin{spliteq}
\bra{0} \mathcal{T} Q(\eta, \vec{x}) Q(\eta', \vec{x}') &\ket{0} \equiv \Theta(\eta-\eta') \bra{0} Q(\eta, \vec{x}) Q(\eta', \vec{x}') \ket{0} + \Theta(\eta'-\eta) \bra{0} Q(\eta', \vec{x}') Q(\eta, \vec{x}) \ket{0} \\
&= \int \left[ \Theta(\eta-\eta') q_{\vec{p}}(\eta) q^*_{\vec{p}}(\eta') + \Theta(\eta'-\eta) q_{-\vec{p}}(\eta') q^*_{-\vec{p}}(\eta) \right] \mathe^{\mathi \vec{p} (\vec{x}-\vec{x}')} \frac{\total^{n-1} p}{(2\pi)^{n-1}} \eqend{,}
\end{spliteq}
and similarly for the tensor perturbation.

\subsection{Weyl tensor correlation function}

Since $H_{ij}$ and $Q$ are independent at tree level, the tree-level correlation function of the Weyl tensor factorises. Using the expressions~\eqref{weyl_dyn} for the Weyl tensor in terms of the dynamical variables and the solutions~\eqref{hkl_corr} and \eqref{q_corr}, we obtain
\begin{equations}
\begin{split}
\bra{0} C^{0i}{}_{0k}(x) C^{0j}{}_{0l}(x') \ket{0} &= \frac{1}{4 (n-2)^2} \left( (n-3) \partial_\eta^2 + \laplace^\vec{x} \right) \left( (n-3) \partial_{\eta'}^2 + \laplace^{\vec{x}'} \right) \bra{0} H^i_k(x) H^j_l(x') \ket{0} \\
&\quad+ \frac{(n-3)^2}{4 (n-1)^2 (n-2)^2} \Pi^i_k(x) \Pi^j_l(x') \mathcal{Q}(x,x') \eqend{,}
\end{split} \\
\bra{0} C^{0i}{}_{0k}(x) C^{0j}{}_{mn}(x') \ket{0} &= \frac{1}{2(n-2)} \left( (n-3) \partial_\eta^2 + \laplace^\vec{x} \right) \partial_{\eta'} \partial^{\vec{x}'}_{[m} \bra{0} H^i_k(x) H_{n]}^j{}(x') \ket{0} \eqend{,} \\
\begin{split}
\bra{0} C^{0i}{}_{0k}(x) C^{mn}{}_{pq}(x') \ket{0} &= \frac{n-3}{(n-1)^2 (n-2)^2} \Pi^i_k(x) \Pi^{[m}_{[p}(x') \delta^{n]}_{q]} \mathcal{Q}(x,x') \\
&\hspace{-6em}+ \frac{1}{(n-2)^2} \left( (n-3) \partial_\eta^2 + \laplace^\vec{x} \right) \left( (n-2) \partial^{[m}_{\vec{x}'} \partial_{[p}^{\vec{x}'} - \delta^{[m}_{[p} \partial^2_{x'} \right) \bra{0} H^i_k(x) H^{n]}_{q]}(x') \ket{0} \eqend{,}
\end{split} \\
\bra{0} C^{0i}{}_{kl}(x) C^{0j}{}_{mn}(x') \ket{0} &= \partial^\vec{x}_{[k} \partial^{\vec{x}'}_{[m} \partial_\eta \partial_{\eta'} \bra{0} H_{l]}^i{}(x) H_{n]}^j{}(x') \ket{0} \eqend{,} \\
\bra{0} C^{0i}{}_{kl}(x) C^{mn}{}_{pq}(x') \ket{0} &= \frac{2}{n-2} \partial_\eta \partial_{[k}^\vec{x} \left( - (n-2) \partial^{[m}_{\vec{x}'} \partial_{[p}^{\vec{x}'} + \delta^{[m}_{[p} \partial^2_{x'} \right) \bra{0} H_{l]}^i(x) H^{n]}_{q]}(x') \ket{0} \eqend{,} \\
\begin{split}
\bra{0} C^{ij}{}_{kl}(x) C^{mn}{}_{pq}(x') \ket{0} &= \frac{4}{(n-1)^2 (n-2)^2} \Pi^{[i}_{[k}(x) \delta^{j]}_{l]} \Pi^{[m}_{[p}(x') \delta^{n]}_{q]} \mathcal{Q}(x,x') \\
&\hspace{-6em}+ \frac{4}{(n-2)^2} \left( (n-2) \partial^{[i}_\vec{x} \partial_{[k}^\vec{x} - \delta^{[i}_{[k} \partial^2_x \right) \left( (n-2) \partial^{[m}_{\vec{x}'} \partial_{[p}^{\vec{x}'} - \delta^{[m}_{[p} \partial^2_{x'} \right) \bra{0} H^{j]}_{l]}(x) H^{n]}_{q]}(x') \ket{0} \eqend{,}
\end{split}
\end{equations}
where the function $\mathcal{Q}$ is defined by
\begin{equation}
\mathcal{Q}(x,x') \equiv \bra{0} \left[ \left[ \epsilon(\eta) Q'(x) \right]' - \epsilon(\eta) \laplace Q(x) \right] \left[ \left[ \epsilon(\eta') Q'(x') \right]' - \epsilon(\eta') \laplace Q(x') \right] \ket{0} \eqend{.}
\end{equation}
For the time-ordered correlation function, one just has to replace the correlation functions of the tensor perturbation and the Sasaki-Mukhanov variable by their time-ordered versions. In this case, the temporal derivatives may introduce $\delta$ distributions when they act to the Heaviside $\Theta$ functions.

These general expressions may be vastly simplified by a number of reasonable assumptions. First, let us concentrate only on the Wightman correlation function. In this case, we can apply the equations of motion~\eqref{dyn_eom} for the dynamical perturbations inside the correlation functions. Second, we assume that the mode functions only depend on the magnitude of the wave vector $\abs{\vec{p}}$, such that the quantum state is rotationally symmetric. This assumption is fulfilled for the solution which gives in the appropriate limit the standard Minkowski mode functions, and permits us to perform the angular integrals. Lastly, since no UV divergences arise at tree level, we can restrict to four dimensions. It is then convenient to decompose the Weyl tensor into its electric and magnetic parts~\cite{kramerstephani}, defining
\begin{equations}[weyl_elmag]
E^k_l &\equiv - C^{0k}{}_{0l} \eqend{,} \\
B^k_l &\equiv \frac{1}{2} \epsilon_{ijl} C^{0k}{}_{ij}
\end{equations}
with the three-dimensional Levi-Civita symbol $\epsilon_{ijl}$. Note that both tensors are symmetric and traceless, which for the magnetic part can be seen by using the Weyl tensor's antisymmetry and tracelessness. 
From these parts, the Weyl tensor can be reconstructed according to
\begin{equations}[weyl_reconstruct]
C^{0j}{}_{0l} &= - E^j_l \eqend{,} \\
C^{0j}{}_{kl} &= - \epsilon_{klm} B^j_m \eqend{,} \\
C^{ij}{}_{kl} &= 4 \delta^{[i}_{[k} E_{l]}^{j]} \eqend{.}
\end{equations}
This reconstruction is possible because of the Weyl tensor symmetries, which in four dimensions completely determine its purely spatial part from the mixed components. 
There are thus only three independent correlation functions, which after the above simplifications read
\begin{equations}[elmag_corr]
\begin{split}
\bra{0} E_{ij}(x) E_{kl}(x') \ket{0} &= \frac{1}{4 \pi^2} \left[ \left( P_{k(i} \laplace \right) \left( P_{j)l} \laplace \right) - \frac{1}{2} \left( P_{kl} \laplace \right) \left( P_{ij} \laplace \right) \right] \\
&\qquad\times \int_0^\infty \left[ H a f_p' + p^2 f_p \right](\eta) \left[ H a f^{*\prime}_p + p^2 f^*_p \right](\eta') \frac{\sin(p r)}{p^3 r} \total p \\
&\quad+ \frac{1}{72 \pi^2} \left[ \epsilon H a \left( 1-\epsilon \right) \right](\eta) \left[ \epsilon H a \left( 1-\epsilon \right) \right](\eta') \\
&\qquad\times \left( \delta_{ij} \laplace - 3 \partial_i \partial_j \right) \left( \delta_{kl} \laplace - 3 \partial_k \partial_l \right) \int_0^\infty q'_p(\eta) q^{*\prime}_p(\eta') \frac{\sin(p r)}{p^3 r} \total p \eqend{,}
\end{split} \\
\begin{split}
\bra{0} E_{ij}(x) B_{kl}(x') \ket{0} &= - \frac{1}{8 \pi^2} \left( \epsilon_{ml(i} P_{j)k} \laplace + \epsilon_{mk(i} P_{j)l} \laplace \right) \partial_m \\
&\qquad\times \int_0^\infty \left( H a f_p'(\eta) + p^2 f_p(\eta) \right) f^{*\prime}_p(\eta') \frac{\sin(p r)}{p r} \total p \eqend{,}
\end{split} \\
\begin{split}
\bra{0} B_{ij}(x) B_{kl}(x') \ket{0} &= \frac{1}{8 \pi^2} \Big[ \left( \delta_{ik} \delta_{jl} + \delta_{jk} \delta_{il} - \delta_{ij} \delta_{kl} \right) \laplace^2 + \left( \delta_{kl} \partial_i \partial_j + \delta_{ij} \partial_k \partial_l \right) \laplace \\
&\qquad\quad- \left( \delta_{ik} \partial_j \partial_l + \delta_{il} \partial_j \partial_k + \delta_{jk} \partial_i \partial_l + \delta_{jl} \partial_i \partial_k \right) \laplace \\
&\qquad\quad+ \partial_k \partial_l \partial_i \partial_j \Big] \int_0^\infty f'_p(\eta) f^{*\prime}_p(\eta') \frac{\sin(p r)}{p r} \total p \eqend{.}
\end{split}
\end{equations}
To present the result in a nice short form, we have introduced the abbreviations $p = \abs{\vec{p}}$ and $r = \abs{\vec{x} - \vec{x}'}$.

\section{Slow roll}
\label{sec_slowroll}

In this section, we again restrict to four dimensions. Apart from the already introduced deceleration parameter $\epsilon$~\eqref{eps_def}, we also define
\begin{equation}
\delta \equiv \frac{\epsilon'}{2 H a \epsilon} = \frac{H'' H - 2 (H')^2 - H' H^2 a}{2 H' H^2 a} \eqend{.}
\end{equation}
We then work to first order in the slow-roll parameters $\epsilon$ and $\delta$. By the definition of $\delta$, the time derivative of $\epsilon$ is of order $\bigo{\delta \epsilon}$, and can be neglected. We assume the same to hold for the time derivative of $\delta$.

The equations of motion for the dynamical perturbations~\eqref{dyn_eom}, expressed for the mode functions~\eqref{modefunctions} can then be written as
\begin{equations}
f_p'' + 2 H a f_p' + p^2 f_p &= 0 \eqend{,} \\
q_p'' + 2 (1+\delta) H a q_p' + p^2 q_p &= 0 \eqend{.}
\end{equations}
To first order in slow-roll, we can change the time derivatives to $x$-derivatives, with $x$ defined as
\begin{equation}
x \equiv \frac{p}{H a (1-\epsilon)} \eqend{,}
\end{equation}
via
\begin{equation}
\partial_\eta = x' \partial_x = - p \partial_x + \bigo{\epsilon^2, \epsilon \delta} \eqend{.}
\end{equation}
Then the equations of motion reduce to
\begin{equations}
\partial_x f_p^2 - \frac{2+2\epsilon}{x} \partial_x f_p + f_p &= \bigo{\epsilon^2, \epsilon \delta} \eqend{,} \\
\partial_x^2 q_p - \frac{2+2\delta+2\epsilon}{x} \partial_x q_p + q_p &= \bigo{\epsilon^2, \epsilon \delta} \eqend{,}
\end{equations}
and correctly normalised~\eqref{normalisation} solutions are given by
\begin{equations}[slowroll_modef]
f_p &= \kappa \sqrt{\frac{\pi}{4 H a^3 (1-\epsilon)}} \hankel1_{\frac{3}{2}+\epsilon}\left( \frac{p}{H a (1-\epsilon)} \right) \eqend{,} \\
q_p &= \kappa \sqrt{\frac{\pi}{4 \epsilon H a^3 (1-\epsilon)}} \hankel1_{\frac{3}{2}+\delta+\epsilon}\left( \frac{p}{H a (1-\epsilon)} \right) \eqend{.}
\end{equations}
These solutions reduce to positive-frequency exponentials in the Minkowski limit $\epsilon,\delta \to 0$, $H \to 0$, $a \to 1$, and define the so-called Bunch-Davies vacuum state.

To calculate the integrals over $p$ appearing in the correlation functions~\eqref{hkl_corr} and~\eqref{q_corr} (and later on in~\eqref{elmag_corr}), we insert a convergence factor $\exp(-\mu p)$ (which gives the correct $\mathi 0$ prescription for the Wightman functions) and a lower (infrared) cutoff $\xi$. The results will generally be divergent as $\xi \to 0$ as can be seen by the behaviour of the integrand near $p = 0$. These infrared divergences are physical: they show that the object that is studied is physically not well-defined and that something else must be computed to be able to relate to measurements. We will later see that in this case the proper observable is the Weyl tensor correlation function~\eqref{elmag_corr}.

Let us thus start with the only scalar integral in~\eqref{elmag_corr}, which to first order in slow-roll (together with the numerical prefactor, but without the spatial derivatives) reads
\begin{spliteq}
I_1 &= \frac{1}{72 \pi^2} \left[ \epsilon H a \left( 1-\epsilon \right) \right](\eta) \left[ \epsilon H a \left( 1-\epsilon \right) \right](\eta') \int_\xi^\infty \mathe^{-\mu p} q'_p(\eta) q^{*\prime}_p(\eta') \frac{\sin(p r)}{p^3 r} \total p \\
&= \frac{\kappa^2}{288 \pi} \epsilon \sqrt{\frac{H(\eta) H(\eta')}{a(\eta) a(\eta')}} \int_\xi^\infty \mathe^{-\mu p} \hankel1_{\frac{1}{2}}\left( \frac{p}{H a}(\eta) \right) \hankel2_{\frac{1}{2}}\left( \frac{p}{H a}(\eta') \right) \frac{\sin(p r)}{p^3 r} \total p \eqend{.}
\end{spliteq}
Because of the overall factor $\epsilon$, the Hankel functions simplify and we can perform the integral easily. Furthermore, we can use the de Sitter expression
\begin{equation}
a = - \frac{1}{H \eta}
\end{equation}
for the scale factor with constant $H$, since any deviations are of higher order in the slow-roll parameters. Then integral then reduces to
\begin{equation}
I_1 = \frac{\kappa^2 \epsilon H^2}{144 \pi^2 r} \int_\xi^\infty \mathe^{-\mu p - \mathi p (\eta-\eta')} p^{-4} \sin(p r) \total p \eqend{,}
\end{equation}
and in the limit $\mu, \xi \to 0$ we obtain $I_1 = I_1^\text{IR} + I_1^\text{fin} + \bigo{\xi}$ with
\begin{equations}[i1_result]
I_1^\text{IR} &= \frac{\kappa^2 \epsilon H^2}{288 \pi^2} \left[ \frac{1}{\xi^2} - \frac{2 \mathi (\eta-\eta')}{\xi} + \frac{r^2 + 3 (\eta-\eta')^2}{18} (6 \gamma - 11 + 6 \ln \xi) \right] \eqend{,} \label{i1_ir} \\
I_1^\text{fin} &= \frac{\kappa^2 \epsilon H^2}{1728 \pi^2 r} \left[ (r + \eta-\eta')^3 \ln \left( \mathi (r + \eta-\eta') \right) + (r - \eta+\eta')^3 \ln \left( - \mathi (r - \eta+\eta') \right) \right] \eqend{.}
\end{equations}
However, there are still spatial derivatives acting on this result, and one calculates
\begin{equation}
\left( \delta_{kl} \laplace - 3 \partial_k \partial_l \right) I_1 = \left( \delta_{kl} - 3 \frac{r_k r_l}{r^2} \right) \left( \partial_r^2 - \frac{1}{r} \partial_r \right) I_1 \eqend{.}
\end{equation}
It is easy to see that $I_1^\text{IR}$~\eqref{i1_ir} is annihilated by these derivatives, so that the scalar contribution to the Weyl tensor is infrared convergent.

For the tensor integrals, we obtain similarly
\begin{equations}[ik_expr]
\begin{split}
I_2 &= \frac{1}{4 \pi^2} \int_\xi^\infty \mathe^{-\mu p} \left[ H a f_p' + p^2 f_p \right](\eta) \left[ H a f^{*\prime}_p + p^2 f^*_p \right](\eta') \frac{\sin(p r)}{p^3 r} \total p \\
&= \frac{\kappa^2}{8 \pi^2} (1+\epsilon) \left[ (Ha)(\eta) \partial_\eta - \laplace \right] \left[ (Ha)(\eta') \partial_{\eta'} - \laplace \right] \bigg[ H(\eta) H(\eta') J \bigg] \eqend{,}
\end{split} \\
\begin{split}
I_3 &= - \frac{1}{8 \pi^2} \int_\xi^\infty \mathe^{-\mu p} \left( H a f_p'(\eta) + p^2 f_p(\eta) \right) f^{*\prime}_p(\eta') \frac{\sin(p r)}{p r} \total p \\
&= \frac{\kappa^2}{16 \pi^2} (1+\epsilon) \left[ (Ha)(\eta) \partial_\eta - \laplace \right] \partial_{\eta'} \laplace \bigg[ H(\eta) H(\eta') J \bigg]
\end{split} \\
I_4 &= \frac{1}{8 \pi^2} \int_\xi^\infty \mathe^{-\mu p} f'_p(\eta) f^{*\prime}_p(\eta') \frac{\sin(p r)}{p r} \total p = - \frac{\kappa^2}{16 \pi^2} (1+\epsilon) \partial_\eta \partial_{\eta'} \laplace \bigg[ H(\eta) H(\eta') J \bigg] \eqend{,}
\end{equations}
with the common integral
\begin{equation}
J = \frac{\pi}{2} (\tau\tau')^\frac{3}{2} \int_\xi^\infty \mathe^{-\mu p} \hankel1_{\frac{3}{2}+\epsilon}\left( - p \tau (1+\epsilon) \right) \hankel2_{\frac{3}{2}+\epsilon}\left( - p \tau' (1+\epsilon) \right) \frac{\sin(p r)}{p^3 r} \total p \eqend{,}
\end{equation}
where we defined
\begin{equation}
\label{tau_def}
\tau \equiv - \frac{1}{H(\eta) a(\eta)} \eqend{,} \qquad \tau' \equiv - \frac{1}{H(\eta') a(\eta')}
\end{equation}
to shorten this and the following expressions. This last integral is more involved, and we need to expand the Hankel functions in $\epsilon$ to get a tractable expression. The derivatives of Hankel functions with respect to their order were calculated in~\cite{brychkovgeddes2005}, and we have~\cite{froebetal2013}
\begin{spliteq}
\label{hankel_expansion}
\sqrt{\frac{\pi}{2}} z^{\frac{3}{2}+\epsilon} \hankel1_{\frac{3}{2}+\epsilon}(z) = - (z+\mathi) \mathe^{\mathi z} - \epsilon &\bigg[ 2 \mathi \mathe^{\mathi z} + 2 \left( z \cos z - \sin z \right) \ln (-\mathi z) \\
&\quad+ \left( z - \mathi \right) \mathe^{-\mathi z} \left[ \Ein\left( 2 \mathi z \right) + \gamma + \ln 2 \right] \bigg] + \bigo{\epsilon^2} \eqend{,}
\end{spliteq}
where $\Ein$ is the regular part of the exponential integral defined by
\begin{equation}
\label{ein_def}
\Ein(z) \equiv \int_0^z \frac{\mathe^t - 1}{t} \total t = \sum_{k=1}^\infty \frac{z^k}{k k!} \eqend{.}
\end{equation}
The integral over $p$ can then be done easily\footnote{The integrals involving the $\Ein$ function can be done term-by-term since both the series expansion and the integral converge absolutely.} and we obtain in the limit $\mu, \xi \to 0$ that $J = J^\text{IR} + J^\text{fin} + \bigo{\xi}$ with
\begin{spliteq}
J^\text{IR} &= \frac{\epsilon}{8 \xi^4} + \frac{1}{4 \xi^4} \left[ 1 + \xi^2 \left( \tau^2 + (\tau')^2 - \frac{1}{3} r^2 \right) \right] \left( 1 - 2 \gamma \epsilon - \epsilon \ln(4 \xi^2 \tau \tau') \right) + \frac{\epsilon \ln \xi}{2} (\tau^4 + (\tau')^4) \\
&\quad- \frac{\ln \xi}{120} \left[ r^4 - 10 r^2 (\tau^2 + (\tau')^2) - 15 (\tau^2 - (\tau')^2)^2 \right] \left( 1 + \epsilon - 2 \gamma \epsilon - \epsilon \ln (4 \xi \tau \tau') \right) \eqend{,}
\end{spliteq}
and where $J^\text{fin}$ is a somewhat lengthy finite part independent of the infrared cutoff $\xi$. Let us first see what happens with the cutoff-dependent terms. Inserting $J^\text{IR}$ into the integrals $I_k$~\eqref{ik_expr} and performing the remaining derivatives, we see that
\begin{equation}
I_2^\text{IR} = I_3^\text{IR} = I_4^\text{IR} = 0 \eqend{,}
\end{equation}
so that the IR cutoff does not make any contribution to the Weyl tensor correlation functions.

For the finite part, we obtain
\begin{equations}[ik_fin]
I_2^\text{fin} &= \frac{\kappa^2}{16 \pi^2} H(\eta') H(\eta) \left[ \frac{1}{1-Z} + \frac{\epsilon}{1+Z} \ln \left( \frac{1-Z}{2} \right) + \epsilon \frac{\tau-\tau'}{r} \ln \left( \frac{\tau-\tau'+r}{\tau-\tau'-r} \right) \right] \eqend{,} \\
I_3^\text{fin} &= \frac{\kappa^2}{32 \pi^2} \frac{H(\eta) H(\eta')}{\tau \tau'} \left[ \frac{\tau-\tau'}{(1-Z)^2} + \frac{\epsilon (\tau+\tau')}{(1+Z)^2} \ln \left( \frac{1-Z}{2} \right) + \frac{2 \epsilon (\tau-\tau')}{(1-Z)^2 (1+Z)} + \frac{\epsilon \tau'}{1+Z} \right] \eqend{,} \\
I_4^\text{fin} &= \frac{\kappa^2}{32 \pi^2} H(\eta) H(\eta') \left[ \frac{1}{1-Z} - \frac{\epsilon}{1+Z} \ln \left( \frac{1-Z}{2} \right) \right] \eqend{,}
\end{equations}
where we defined
\begin{equation}
\label{z_def}
Z \equiv 1 - \frac{r^2 - (1+2\epsilon) (\tau-\tau')^2}{2 \tau \tau'} \eqend{.}
\end{equation}

To obtain the final result, it only remains to perform some spatial derivatives, and we obtain from the expressions~\eqref{elmag_corr}
\begin{spliteq}
\label{weyl_corr_ee}
\bra{0} E_{ij}(x) E_{kl}(x') \ket{0} &= \left[ \left( P_{k(i} \laplace \right) \left( P_{j)l} \laplace \right) - \frac{1}{2} \left( P_{kl} \laplace \right) \left( P_{ij} \laplace \right) \right] I_2^\text{fin} \\
&\qquad+ \left( \delta_{ij} \laplace - 3 \partial_i \partial_j \right) \left( \delta_{kl} \laplace - 3 \partial_k \partial_l \right) I_1^\text{fin} \\
&= \left( \delta_{k(i} \delta_{j)l} - \frac{1}{3} \delta_{ij} \delta_{kl} \right) \left( \partial_r^4 + \frac{2}{r} \partial_r^3 - \frac{3}{r^2} \partial_r^2 + \frac{3}{r^3} \partial_r \right) I_2^\text{fin} \\
&\qquad- \frac{1}{6} R_{ij} R_{kl} \left( \partial_r^4 + \frac{2}{r} \partial_r^3 - \frac{9}{r^2} \partial_r^2 + \frac{9}{r^3} \partial_r \right) I_2^\text{fin} \\
&\qquad- \frac{2}{r^2} r_{(i} \left( \delta_{j)(k} - \frac{r_{j)} r_{(k}}{r^2} \right) r_{l)} \left( \partial_r^4 - \frac{3}{r^2} \partial_r^2 + \frac{3}{r^3} \partial_r \right) I_2^\text{fin} \\
&\quad+ \bigg[ - \frac{6}{r^2} \left( \delta_{ij} \delta_{kl} - 3 \delta_{i(k} \delta_{l)j} \right) + 36 r_{(i} \left( \delta_{j)(k} - \frac{r_j r_k}{r^2} \right) r_{l)} \left( \frac{1}{r^3} \partial_r - \frac{2}{r^4} \right) \\
&\qquad\quad+ R_{ij} R_{kl} \left( \partial_r^2 - \frac{1}{r} \partial_r \right) \bigg] \left( \partial_r^2 - \frac{1}{r} \partial_r \right) I_1^\text{fin} \eqend{,}
\end{spliteq}
\begin{spliteq}
\label{weyl_corr_eb}
\bra{0} E_{ij}(x) B_{kl}(x') \ket{0} &= \left( \epsilon_{ml(i} P_{j)k} \laplace + \epsilon_{mk(i} P_{j)l} \laplace \right) \partial_m I_3^\text{fin} \\
&= \frac{r_m}{(\tau \tau')^3} \epsilon_{mk(i} \left( \delta_{j)l} \left( 4 \tau \tau' \partial_Z^2 I_3^\text{fin} - r^2 \partial_Z^3 I_3^\text{fin} \right) + r_{j)} r_l \partial_Z^3 I_3^\text{fin} \right) \\
&\quad+ \frac{r_m}{(\tau \tau')^3} \epsilon_{ml(i} \left( \delta_{j)k} \left( 4 \tau \tau' \partial_Z^2 I_3^\text{fin} - r^2 \partial_Z^3 I_3^\text{fin} \right) + r_{j)} r_k \partial_Z^3 I_3^\text{fin} \right) \eqend{,}
\end{spliteq}
and
\begin{spliteq}
\label{weyl_corr_bb}
\bra{0} B_{ij}(x) B_{kl}(x') \ket{0} &= \Big[ \left( \delta_{ik} \delta_{jl} + \delta_{jk} \delta_{il} - \delta_{ij} \delta_{kl} \right) \laplace^2 + \left( \delta_{kl} \partial_i \partial_j + \delta_{ij} \partial_k \partial_l \right) \laplace \\
&\qquad\quad- \left( \delta_{ik} \partial_j \partial_l + \delta_{il} \partial_j \partial_k + \delta_{jk} \partial_i \partial_l + \delta_{jl} \partial_i \partial_k \right) \laplace + \partial_k \partial_l \partial_i \partial_j \Big] I_4^\text{fin} \\
&= 2 \left( \delta_{i(k} \delta_{l)j} - \frac{1}{3} \delta_{ij} \delta_{kl} \right) \left[ \frac{6}{(\tau \tau')^2} \partial_Z^2 I_4^\text{fin} - \frac{8 r^2}{(\tau \tau')^3} \partial_Z^3 I_4^\text{fin} + \frac{r^4}{(\tau \tau')^4} \partial_Z^4 I_4^\text{fin} \right] \\
&\quad+ 4 r_{(i} \left( \delta_{j)(k} - \frac{r_{j)} r_{(k}}{r^2} \right) r_{l)} \left[ \frac{6}{(\tau \tau')^3} \partial_Z^3 I_4^\text{fin} - \frac{r^2}{(\tau \tau')^4} \partial_Z^4 I_4^\text{fin} \right] \\
&\quad+ \frac{1}{3} R_{ij} R_{kl} r^2 \left[ \frac{8}{(\tau \tau')^3} \partial_Z^3 I_4^\text{fin} - \frac{r^2}{(\tau \tau')^4} \partial_Z^4 I_4^\text{fin} \right]
\end{spliteq}
with the traceless tensor
\begin{equation}
R_{ij} \equiv \delta_{ij} - 3 \frac{r_i r_j}{r^2} \eqend{.}
\end{equation}
Note that the derivatives in equation~\eqref{weyl_corr_ee} with respect to $r$ do also act on $Z$ through its implicit dependence on $r$ as can be seen from the definition~\eqref{z_def}. We do not perform the remaining $Z$ and $r$ derivatives; they are straightforward to do, but the resulting expressions are lengthy and not particularly enlightening.

\subsection{The de Sitter limit}

In the de Sitter limit, we have $\epsilon = \delta = 0$, the Hubble parameter $H$ becomes a constant, and the de Sitter scale factor is given by
\begin{equation}
a = - \frac{1}{H \eta} \eqend{,}
\end{equation}
such that $\tau$ and $\tau'$~\eqref{tau_def} reduce simply to $\eta$ and $\eta'$. Furthermore, $Z$ as defined in~\eqref{z_def} goes over into the usual de Sitter measure of geodesic distance $\mu$,
\begin{equation}
Z = \cos(H\mu) \eqend{.}
\end{equation}
The Weyl tensor correlator in de Sitter space has been calculated several times~\cite{kouris2001,morawoodard2012,moratsamiswoodard2012a,kouris2001corr,froebrouraverdaguer2014}, but to compare with our results the explicit expressions in~\cite{froebrouraverdaguer2014} are suited best. Explicitly, the result reads (adapted to our notation)
\begin{spliteq}
\label{weyl_corr_ds}
\bra{0} {\tilde{C}^{\alpha\beta}}{}_{\gamma\delta}(x) {\tilde{C}^{\mu'\nu'}}{}_{\rho'\sigma'}(x') \ket{0} &= \frac{\kappa^2 H^6}{2 \pi^2 (1-Z)^5} \bigg[ (1-Z)^2 (1-3Z) {}^{(1)}\mathcal{R} + 6 (1-Z) {}^{(2)}\mathcal{R} \\
&\qquad- 12 (3-Z) {}^{(3)}\mathcal{R} - 6 (2-Z) (1-Z) {}^{(4)}\mathcal{R} - 6 (1-Z)^3 {}^{(5)}\mathcal{R} \\
&\qquad- 6 (1-Z) {}^{(6)}\mathcal{R} + 6 (5-Z) {}^{(7)}\mathcal{R} + 2 (4-Z) (1-Z) {}^{(8)}\mathcal{R} \\
&\qquad+ (3-Z) (1-Z)^2 {}^{(9)}\mathcal{R} \bigg]{}^{[\alpha\beta]}{}_{[\gamma\delta]}{}^{[\mu'\nu']}{}_{[\rho'\sigma']} \eqend{,}
\end{spliteq}
where ${}^{(k)}\mathcal{R}$ are a set of bitensors defined by
\begin{spliteq}
{}^{(1)}\mathcal{R}_{\alpha\beta\gamma\delta\mu'\nu'\rho'\sigma'} &= \tilde{g}^{(0)}_{\alpha\gamma} \tilde{g}^{(0)}_{\beta\delta} \tilde{g}^{(0)}_{\mu'\rho'} \tilde{g}^{(0)}_{\nu'\sigma'} \\
{}^{(2)}\mathcal{R}_{\alpha\beta\gamma\delta\mu'\nu'\rho'\sigma'} &= H^{-2} \tilde{g}^{(0)}_{\alpha\gamma} \tilde{g}^{(0)}_{\mu'\rho'} \left( \tilde{g}^{(0)}_{\beta\delta} Z_{;\nu'} Z_{;\sigma'} + Z_{;\beta} Z_{;\delta} \tilde{g}^{(0)}_{\nu'\sigma'} \right) \\
{}^{(3)}\mathcal{R}_{\alpha\beta\gamma\delta\mu'\nu'\rho'\sigma'} &= H^{-4} \tilde{g}^{(0)}_{\alpha\gamma} \tilde{g}^{(0)}_{\mu'\rho'} Z_{;\beta} Z_{;\delta} Z_{;\nu'} Z_{;\sigma'} \\
{}^{(4)}\mathcal{R}_{\alpha\beta\gamma\delta\mu'\nu'\rho'\sigma'} &= 4 H^{-4} \tilde{g}^{(0)}_{\alpha\gamma} \tilde{g}^{(0)}_{\mu'\rho'} Z_{;(\beta} Z_{;\delta)(\nu'} Z_{;\sigma')} \\
{}^{(5)}\mathcal{R}_{\alpha\beta\gamma\delta\mu'\nu'\rho'\sigma'} &= 2 H^{-4} \tilde{g}^{(0)}_{\alpha\gamma} \tilde{g}^{(0)}_{\mu'\rho'} Z_{;\beta(\nu'} Z_{;\sigma')\delta} \\
{}^{(6)}\mathcal{R}_{\alpha\beta\gamma\delta\mu'\nu'\rho'\sigma'} &= 2 H^{-6} \left( \tilde{g}^{(0)}_{\alpha\gamma} Z_{;\mu'} Z_{;\rho'} + Z_{;\alpha} Z_{;\gamma} \tilde{g}^{(0)}_{\mu'\rho'} \right) Z_{;\beta(\nu'} Z_{;\sigma')\delta} \\
{}^{(7)}\mathcal{R}_{\alpha\beta\gamma\delta\mu'\nu'\rho'\sigma'} &= 2 H^{-8} Z_{;\alpha} Z_{;\gamma} Z_{;\mu'} Z_{;\rho'} Z_{;\beta(\nu'} Z_{;\sigma')\delta} \\
{}^{(8)}\mathcal{R}_{\alpha\beta\gamma\delta\mu'\nu'\rho'\sigma'} &= 8 H^{-8} Z_{;(\alpha} Z_{;\gamma)(\mu'} Z_{;\rho')} Z_{;\beta(\nu'} Z_{;\sigma')\delta} \\
{}^{(9)}\mathcal{R}_{\alpha\beta\gamma\delta\mu'\nu'\rho'\sigma'} &= 4 H^{-8} Z_{;\alpha(\mu'} Z_{;\rho')\gamma} Z_{;\beta(\nu'} Z_{;\sigma')\delta} \eqend{,}
\end{spliteq}
and where a semicolon denotes a covariant derivative with respect to the background metric $\tilde{g}^{(0)}_{\mu\nu}$ (which for the shown bitensors are equivalent to partial derivatives). Furthermore, primed indices indicate that the derivative must be taken at the point $x'$.

From the definition of $Z$~\eqref{z_def}, recalling that in de Sitter space we have $\tau = \eta$, we can easily calculate
\begin{equations}
\partial_\eta Z &= Z_{;0} = \frac{r^2 + \eta^2 - (\eta')^2}{2 \eta^2 \eta'} = \frac{1}{\eta'} - \frac{Z}{\eta} \eqend{,} \\
\partial_{\eta'} Z &= Z_{;0'} = \frac{r^2 - \eta^2 + (\eta')^2}{2 \eta (\eta')^2} = \frac{1}{\eta} - \frac{Z}{\eta'} \eqend{,} \\
\partial^\vec{x}_i Z &= Z_{;i} = - \frac{r_i}{\eta \eta'} \eqend{,} \\
\partial^{\vec{x}'}_i Z &= Z_{;i'} = \frac{r_i}{\eta \eta'} \eqend{,} \\
\partial_\eta \partial_{\eta'} Z &= Z_{;00'} = \frac{Z}{\eta \eta'} - \frac{1}{\eta^2} - \frac{1}{(\eta')^2} \eqend{,} \\
\partial_\eta \partial^{\vec{x}'}_i Z &= Z_{;0i'} = - \frac{r_i}{\eta^2 \eta'} \eqend{,} \\
\partial^\vec{x}_i \partial_{\eta'} Z &= Z_{;i0'} = \frac{r_i}{\eta (\eta')^2} \eqend{,} \\
\partial^\vec{x}_i \partial^{\vec{x}'}_j Z &= Z_{;ij'} = \frac{\delta_{ij}}{\eta \eta'} \eqend{.}
\end{equations}

With the rescaling~\eqref{weyl_tilde}, the definitions~\eqref{weyl_elmag} and the above derivatives of $Z$, we can then calculate the electric and magnetic correlators
\begin{spliteq}
&\bra{0} E_{ij}(x) E_{kl}(x') \ket{0} = \frac{1}{H^4 \eta^2 (\eta')^2} \bra{0} {\tilde{C}^{0i}}{}_{0j}(x) {\tilde{C}^{0'k'}}{}_{0'l'}(x') \ket{0} \\
&= \frac{\kappa^2 H^2}{4 \pi^2 \eta^4 (\eta')^4 (1-Z)^5} \bigg[ \left( 3 \delta_{i(k} \delta_{l)j} - \delta_{ij} \delta_{kl} \right) \left[ 2 (\eta-\eta')^4 + 4 (\eta-\eta')^2 \eta \eta' (1-Z) + [ \eta \eta' (1-Z) ]^2 \right] \\
&\qquad- R_{ij} R_{kl} (\eta-\eta')^2 r^2 - 6 r_{(i} \left( \delta_{j)(k} - \frac{r_{j)} r_{(k}}{r^2} \right) r_{l)} \left( 2 (\eta-\eta')^2 + (1-Z) \eta \eta' \right) \bigg] \eqend{,}
\end{spliteq}
\begin{spliteq}
&\bra{0} E_{ij}(x) B_{kl}(x') \ket{0} = - \frac{1}{2 H^4 \eta^2 (\eta')^2} \epsilon_{lmn} \bra{0} {\tilde{C}^{0i}}{}_{0j}(x) {\tilde{C}^{0'k'}}{}_{m'n'}(x') \ket{0} \\
&= - \frac{3 \kappa^2 H^2 (\eta-\eta')}{4 \pi^2 \eta^4 (\eta')^4 (1-Z)^5} r_m \epsilon_{mk(i} \left[ \delta_{j)l} \left[ (\eta - \eta')^2 + (1-Z) \eta \eta' \right] - r_{j)} r_l \right] \\
&\quad- \frac{3 \kappa^2 H^2 (\eta-\eta')}{4 \pi^2 \eta^4 (\eta')^4 (1-Z)^5} r_m \epsilon_{ml(i} \left[ \delta_{j)k} \left[ (\eta - \eta')^2 + (1-Z) \eta \eta' \right] - r_{j)} r_k \right] \\
\end{spliteq}
and
\begin{spliteq}
&\bra{0} B_{ij}(x) B_{kl}(x') \ket{0} = \frac{\epsilon_{jpq} \epsilon_{lmn}}{4 H^4 \eta^2 (\eta')^2} \bra{0} {\tilde{C}^{0i}}{}_{pq}(x) {\tilde{C}^{0'k'}}{}_{m'n'}(x') \ket{0} \\
&= \frac{\kappa^2 H^2}{4 \eta^4 (\eta')^4 \pi^2 (1-Z)^5} \bigg[ - 6 r_{(i} \left( \delta_{j)(k} - \frac{r_{j)} r_{(k}}{r^2} \right) r_{l)} \left( 2 (\eta-\eta')^2 + (1-Z) \eta \eta' \right) \\
&\qquad+ 3 \left( \delta_{i(k} \delta_{l)j} - \frac{1}{3} \delta_{ij} \delta_{kl} \right) \left( 2 (\eta-\eta')^4 + 4 (\eta-\eta')^2 (1-Z) \eta \eta' + (1-Z)^2 \eta^2 (\eta')^2 \right) \\
&\qquad- R_{ij} R_{kl} r^2 (\eta-\eta)^2 \bigg] \eqend{.}
\end{spliteq}
Comparing these results with the de~Sitter limit of the slow-roll correlation functions~\eqref{weyl_corr_ee}, \eqref{weyl_corr_eb} and \eqref{weyl_corr_bb} (after performing the derivatives with respect to $Z$), we find complete agreement.

\section{Recovering the power spectrum}
\label{sec_powerspectrum}

The usual power spectra are defined from the equal-time limit of the Fourier-transformed two-point function, \ie, for the tensor power spectrum we have
\begin{equation}
\label{tensor_spectrum}
\mathcal{P}_\text{T}(\abs{\vec{k}}, \eta) = \frac{1}{4 (2\pi)^3} \abs{\vec{k}}^3 \delta_{ik} \delta_{jl} \int \bra{0} H_{ij}(\eta, \vec{x}) H_{kl}(\eta, 0) \ket{0} \mathe^{- \mathi \vec{k} \vec{x}} \total^3 x \eqend{.}
\end{equation}
and similarly for the scalar power spectrum
\begin{equation}
\mathcal{P}_\text{S}(\abs{\vec{k}}, \eta) = \frac{1}{(2\pi)^3} \abs{\vec{k}}^3 \int \bra{0} Q(\eta, \vec{x}) Q(\eta, 0) \ket{0} \mathe^{- \mathi \vec{k} \vec{x}} \total^3 x \eqend{.}
\end{equation}
Using the tree level correlation functions~\eqref{hkl_corr} and \eqref{q_corr} in four dimensions, these expressions reduce to the well-known
\begin{equation}
\mathcal{P}_\text{T}(\abs{\vec{k}}, \eta) = \frac{1}{(2\pi)^3} \abs{\vec{k}}^3 f_{\vec{k}}(\eta) f^*_{\vec{k}}(\eta)
\end{equation}
and
\begin{equation}
\mathcal{P}_\text{S}(\abs{\vec{k}}, \eta) = \frac{1}{(2\pi)^3} \abs{\vec{k}}^3 q_{\vec{k}}(\eta) q^*_{\vec{k}}(\eta) \eqend{.}
\end{equation}
For the slow-roll mode functions~\eqref{slowroll_modef}, we can expand the Hankel functions for small momenta
\begin{equation}
\hankel1_\nu(x) \hankel2_\nu(x) \sim \frac{\Gamma^2(\nu)}{\pi^2} \left( \frac{2}{x} \right)^{2\nu}
\end{equation}
to obtain as $\abs{\vec{k}} \to 0$
\begin{equations}
\mathcal{P}_\text{T}(\abs{\vec{k}}, \eta) &\sim \frac{\kappa^2 H^2}{16 \pi^3} \left[ 1 + 2 \epsilon (1-\gamma-2\ln2) \right] \left( \frac{2 H a}{\abs{\vec{k}}} \right)^{2\epsilon} + \bigo{\epsilon^2} \eqend{,} \\
\mathcal{P}_\text{S}(\abs{\vec{k}}, \eta) &\sim \frac{\kappa^2 H^2}{16 \pi^3 \epsilon} \left[ 1 + 2 (\epsilon+\delta) (1-\gamma-2\ln2) + 2 \delta \right] \left( \frac{2 H a}{\abs{\vec{k}}} \right)^{2\delta+2\epsilon} + \bigo{\epsilon,\delta}
\end{equations}
and so the IR problem is even worse than in de Sitter space if $\epsilon > 0$ and/or $\epsilon+\delta > 0$.

What is the relation between the power spectra and the Weyl tensor correlator which, as we have seen in the previous section, is IR finite in slow-roll? From the explicit expression of the Weyl tensor in terms of the gauge-invariant parts of the metric perturbation given in equations~\eqref{weyl_decomp}, one can easily calculate that
\begin{equation}
2 \laplace C^{0k}{}_{0l} - 2 \partial_j \partial_{(k} C^{0j}{}_{0l)} - \partial_j C^{0k}{}_{jl}' - \partial_j C^{0l}{}_{jk}' - 2 \partial_i \partial_j C^{ik}{}_{jl} = \laplace^2 H_{kl} \eqend{.}
\end{equation}
In four dimensions, we can write the Weyl tensor in terms of its electric and magnetic parts~\eqref{weyl_reconstruct} to obtain
\begin{equation}
\label{laplace2_hkl}
\laplace^2 H_{kl} = - 2 \left( 2 \delta_{km} \delta_{ln} \laplace - 3 \partial_m \delta_{n(k} \partial_{l)} + \delta_{kl} \partial_m \partial_n \right) E_{mn} + \epsilon_{kmn} \partial_n B_{ml}' + \epsilon_{lmn} \partial_n B_{mk}' \eqend{.}
\end{equation}
With the explicit expressions~\eqref{elmag_corr} for the electric and magnetic correlators and using the equations of motion~\eqref{dyn_eom} applied to the mode functions~\eqref{normalisation}, we then obtain
\begin{equation}
\label{tensor_2pf}
\bra{0} H_{ij}(x) H_{kl}(x') \ket{0} = \frac{1}{2\pi^2} \left[ 2 P_{k(i} P_{j)l} - P_{ij} P_{kl} \right] \int_0^\infty f_p(\eta) f^*_p(\eta') \frac{\sin(p r)}{r} p \total p \eqend{,}
\end{equation}
which is exactly the result~\eqref{hkl_corr}, in four dimensions and integrated over the angles. This integral thus also yields the well-known nearly scale-invariant power spectrum, and is IR-divergent (even after applying the projection tensors $P_{ij}$), unlike the Weyl tensor correlation function. We can trace the difference back to equation~\eqref{laplace2_hkl}, where we have to invert the squared Laplace operator to obtain the metric perturbation from the Weyl tensor. Denoting the right-hand side of equation~\eqref{laplace2_hkl} by $J_{kl}$, this inversion reads explicitly (but formally)
\begin{equation}
H_{kl}(\eta, \vec{x}) = \frac{1}{16\pi^2} \int \frac{1}{\abs{\vec{x}-\vec{x}'}} \left[ \int \frac{J_{kl}(\eta, \vec{x}'')}{\abs{\vec{x}'-\vec{x}''}} \total^3 x'' \right] \total^3 x' \eqend{.}
\end{equation}
Even for a source $J_{kl}$ of compact spatial support, the first integral (in brackets) has a monopole term, which for large $\abs{x'}$ dominates and only falls off as $\sim 1/\abs{x'}$, and the second integral therefore does not converge. One can of course formally invert the squared Laplace operator in Fourier space, which reads
\begin{equation}
H_{kl}(\eta, \vec{x}) = \int \frac{1}{\abs{\vec{p}}^4} \left[ \int J_{kl}(\eta, \vec{x}') \mathe^{-\mathi \vec{p} \vec{x}'} \total^3 x' \right] \mathe^{\mathi \vec{p} \vec{x}} \frac{\total^3 p}{(2\pi)^3} \eqend{,}
\end{equation}
but then an IR cutoff needs to be introduced to avoid the manifest divergence for small $\abs{\vec{p}}$ --- this IR cutoff is exactly the one that has been used before. It is thus seen that the tensor two-point function can be directly obtained from the Weyl tensor correlation function by using the expression~\eqref{laplace2_hkl}, applying the differential operators to the electric and magnetic correlation functions given by~\eqref{elmag_corr}. However, the inversion of the squared Laplace operator that needs to be done is singular in the IR, and to complete the formal inversion an IR cutoff needs to be introduced. The reason why one has to inverse an elliptic operator and impose boundary conditions is clear: the definition of transversality and tracelessness for the physical graviton is inherently nonlocal, as can be seen, \eg, from the explicit expression for the transverse projector~\eqref{proj_def}. To recover then the usual definition of the power spectrum one (formally) imposes vanishing boundary conditions at infinity, and this leads directly to the usual IR cutoff. Especially, for the tensor power spectrum~\eqref{tensor_spectrum} we obtain
\begin{equation}
\label{power_tensor}
\mathcal{P}_\text{T}(\abs{\vec{k}}, \eta) = \frac{1}{32 \pi^3 \abs{\vec{k}}^5} \int J_{kl}(\eta, \vec{y}) J_{kl}(\eta, 0) \mathe^{-\mathi \vec{k} \vec{y}} \total^3 y \eqend{.}
\end{equation}

It may seem that this procedure does not present any advantage over a direct calculation (with IR cutoff) of the tensor correlation function (or the power spectrum), since we have recovered exactly the result~\eqref{tensor_2pf} of the direct calculation. However, the Weyl tensor is a gauge-invariant observable and it may be easier to calculate it in a different way, \eg, in average gauges and without decomposing the metric perturbations. Our results show that the usual tensor correlation function, including the IR cutoff in position space, can be obtained in a straightforward way from the Weyl correlation function, without losing any information as had been claimed by some authors~\cite{morawoodard2012}. Especially, the rejection of the de Sitter-invariant propagator constructed in~\cite{morrison2013} on the basis that it does not have the correct equal-time limit~\cite{miaoetal2013} can now be seen to be invalid, since this propagator yields the exactly same Weyl tensor correlator~\eqref{weyl_corr_ds}.\footnote{However, the recovery of the dS-invariant propagator (or any other propagator) from the gauge-invariant Weyl tensor is clearly impossible without further input since propagators are gauge-dependent.} In the cases where the Weyl tensor correlator has been computed, this is also the case for all other propagators that had been previously derived in the literature in diverse parts of de Sitter space~\cite{allen1987,allenturyn1987,turyn1990,antoniadismottolo1991,tsamiswoodard1994,devegaramirezsanchez1999,hawkinghertogturok2000,higuchikouris2001,higuchiweeks2003,miaotsamiswoodard2011,moratsamiswoodard2012b,bernarcrispinohiguchi2014}. It is thus seen that the choice of propagator is completely immaterial if one considers gauge-invariant local observables. A prime example where gauge invariance is manifest is the linearised Weyl tensor correlator, and the formula~\eqref{power_tensor} shows that also the power spectrum can be defined in a gauge-invariant way, which also clearly exhibits its nonlocality.\footnote{A similar gauge-invariant definition was given in~\cite{higuchimarolfmorrison2011} in terms of invariant creation and annihilation operators. The two definitions can probably shown to be equal.} Care is needed when taking into account graviton self-interactions and internal graviton loops, and showing propagator equivalence in these cases is an open problem.
The inversion of the squared Laplace operator is highly non-local, but is has been shown in de Sitter spacetime that a local probe of the metric perturbations in the form $\int f^{\mu\nu} h_{\mu\nu} \total^4 x$ with a compactly supported $f^{\mu\nu}$ (subject to conditions that make the integral gauge-invariant) can be rewritten as $\int f^{\mu\nu\rho\sigma} C_{\mu\nu\rho\sigma} \total^4 x$, where $f^{\mu\nu\rho\sigma}$ is locally constructed from $f^{\mu\nu}$, \ie, with compact support arbitrarily close to the support of $f^{\mu\nu}$~\cite{higuchi2012}. This shows that in de Sitter space linearised metric perturbations and Weyl tensor perturbations contain locally the same information, and it is highly probable that a similar proof can be found for FLRW spacetimes. In a similar way, one can reconstruct the scalar power spectrum from the correlator of the spatial curvature scalar.

\section{Conclusion}

In this article, we have derived an explicit formula for the Weyl tensor correlation function in Friedmann-Lema{\^\i}tre-Robertson-Walker spacetimes. The decomposition of the metric perturbations into gauge-invariant and gauge-dependent parts as well as the expression of the Weyl tensor using this decomposition are completely general, and can be used in any metric theory of gravity. For definiteness, we focused on a standard single-field inflationary model in Einstein gravity. The action was then expressed solely in terms of the gauge-invariant parts of the metric and scalar field perturbations, as was expected by its diffeomorphism invariance (which for the perturbations translates into gauge invariance). The constrained degrees of freedom were then eliminated from the action and the Weyl tensor using their equations of motion, and an explicit formula for the Weyl tensor correlation functions in terms of the mode functions for the dynamical perturbations was derived.

To illustrate these general derivations, we quantized the perturbations for the case of slow-roll single-field inflation, and obtained the Weyl tensor correlation function to first order in the slow-roll parameters. Unlike the tensor and scalar correlation functions, the Weyl tensor correlator does not show any IR divergence in this slow-roll case, and it is thus a proper observable. Nevertheless, this does not imply that information is lost by going from the correlation function of the metric perturbations to the Weyl tensor correlator: one can recover the full tensor two-point function from the Weyl tensor by taking derivatives and inverting a squared Laplace operator. The inversion can only be done by introducing a IR cutoff, and in this way one also recovers the well-known IR divergences of the tensor correlator, and the nearly scale invariant power spectrum this implies. Especially, this shows that (at least on the linearised level) any graviton propagator that gives the same Weyl tensor correlation function can be admitted, whether obtained in reduced phase space, through quantisation in average gauges or in any other way. For spacetimes which are expanding more rapidly, \eg, for constant but large deceleration parameter $\epsilon$~\cite{janssenmiaoprokopecwoodard2008}, IR divergent terms may still show up in the Weyl correlator. These cases merit further investigation.

\acknowledgments

It is a pleasure to thank Atsushi Higuchi, Ian Morrison, Takahiro Tanaka, Yuko Urakawa, Enric Verdaguer and Richard Woodard for very useful discussions, comments and references.

Financial support through a FPU scholarship no. AP2010-5453, and partial financial support by the Research Projects MCI FPA2007-66665-C02-02, FPA2010-20807-C02-02, CPAN CSD2007-00042, with\-in the program Consolider-Ingenio 2010, and AGAUR 2009-SGR-00168, is acknowledged, as well as support through ERC starting grant QC\&C 259562.

\appendix

\section{Perturbative expansions}
\label{app_expansion}

In this appendix, we give explicit formulae for the expansion of geometrical quantities to second order in metric perturbations. The metric is decomposed as
\begin{equation}
\tilde{g}_{\mu\nu} = a^2 ( \eta_{\mu\nu} + h_{\mu\nu} ) \eqend{,}
\end{equation}
and the inverse metric reads to second order
\begin{equation}
\tilde{g}^{\mu\nu} = a^{-2} \left( \eta^{\mu\nu} - h^{\mu\nu} + h^{\mu\alpha} h_\alpha^\nu \right) + \bigo{h^3} \eqend{,}
\end{equation}
where the indices on $h_{\mu\nu}$ are raised with the flat Minkowski metric $\eta^{\mu\nu}$, and we define $h \equiv \eta^{\mu\nu} h_{\mu\nu}$. The metric determinant is given by
\begin{equation}
\sqrt{-\tilde{g}} = a^n \left[ 1 + \frac{1}{2} h - \frac{1}{8} \left( 2 h_{\mu\nu} h^{\mu\nu} - h^2 \right) \right] + \bigo{h^3} \eqend{,}
\end{equation}
and the Christoffel symbols read
\begin{spliteq}
\tilde{\Gamma}^\alpha_{\beta\gamma} &= H a \left[ \delta^\alpha_\gamma \delta^0_\beta + \delta^\alpha_\beta \delta^0_\gamma + \left( \delta^\alpha_0 + h^{\alpha0} + h^{\alpha\mu} h_{0\mu} \right) \eta_{\beta\gamma} + \left( \delta^\alpha_0 + h^{\alpha0} \right) h_{\beta\gamma} \right] \\
&\quad+ \frac{1}{2} \left( \eta^{\alpha\mu} - h^{\alpha\mu} \right) \left( \partial_\beta h_{\gamma\mu} + \partial_\gamma h_{\beta\mu} - \partial_\mu h_{\beta\gamma} \right) + \bigo{h^3} \eqend{.}
\end{spliteq}
The Riemann curvature tensor then follows as
\begin{spliteq}
\tilde{R}^{\mu\nu}{}_{\rho\sigma} &= 2 H^2 \delta^\mu_{[\rho} \delta_{\sigma]}^\nu \left( 1 + h_{00} + h^{0\alpha} h_{0\alpha} \right) - 4 \frac{H'}{a} \left( \delta^{[\mu}_0 + h^{0[\mu} + h_{0\alpha} h^{\alpha[\mu} \right) \delta^{\nu]}_{[\rho} \delta^0_{\sigma]} \\
&\quad- \frac{H}{a} \delta^{[\mu}_\rho \left( \partial^{\nu]} h_{0\sigma} + \partial_\sigma h^{\nu]}_0 - \partial_0 h^{\nu]}_\sigma \right) + \frac{H}{a} \delta^{[\mu}_\sigma \left( \partial^{\nu]} h_{0\rho} + \partial_\rho h^{\nu]}_0 - \partial_0 h^{\nu]}_\rho \right) \\
&\quad- \frac{H}{a} h^{0\alpha} \delta^{[\mu}_\rho \left( \partial^{\nu]} h_{\alpha\sigma} + \partial_\sigma h^{\nu]}_\alpha - \partial_\alpha h^{\nu]}_\sigma \right) + \frac{H}{a} h^{0\alpha} \delta^{[\mu}_\sigma \left( \partial^{\nu]} h_{\rho\alpha} + \partial_\rho h^{\nu]}_\alpha - \partial_\alpha h^{\nu]}_\rho \right) \\
&\quad+ \frac{H}{a} \delta^{[\mu}_\rho h^{\nu]\alpha} \left( \partial_\alpha h_{0\sigma} + \partial_\sigma h_{0\alpha} - \partial_0 h_{\alpha\sigma} \right) - \frac{H}{a} \delta^{[\mu}_\sigma h^{\nu]\alpha} \left( \partial_\alpha h_{0\rho} + \partial_\rho h_{0\alpha} - \partial_0 h_{\alpha\rho} \right) \\
&\quad+ \frac{1}{2} a^{-2} \left[ \left( \partial_{[\sigma} h^{\mu\alpha} \right) \partial_{\rho]} h^\nu_\alpha + \left( \partial^{[\nu} h_{\alpha\rho} - \partial_\alpha h^{[\nu}_\rho \right) \partial_\sigma h^{\mu]\alpha} - \left( \partial^{[\nu} h_{\alpha\sigma} - \partial_\alpha h^{[\nu}_\sigma \right) \partial_\rho h^{\mu]\alpha} \right] \\
&\quad+ \frac{1}{4} a^{-2} \left[ \left( \partial^\alpha h^\mu_\rho - \partial^\mu h^\alpha_\rho \right) \left( \partial^\nu h_{\alpha\sigma} - \partial_\alpha h^\nu_\sigma \right) - \left( \partial^\alpha h^\mu_\sigma - \partial^\mu h^\alpha_\sigma \right) \left( \partial^\nu h_{\alpha\rho} - \partial_\alpha h^\nu_\rho \right) \right] \\
&\quad- 2 a^{-2} \partial_{[\rho} \partial^{[\mu} h^{\nu]}_{\sigma]} - 2 a^{-2} h^{\alpha[\mu} \left( \partial^{\nu]} \partial_{[\rho} h_{\sigma]\alpha} - \partial_\alpha \partial_{[\rho} h^{\nu]}_{\sigma]} \right) + \bigo{h^3} \eqend{,}
\end{spliteq}
and the Ricci scalar which we need in the action has the expansion
\begin{spliteq}
\tilde{R} &= n (n-1) H^2 \left( 1 + h_{00} + h^{0\alpha} h_{0\alpha} \right) + 2 (n-1) \frac{H'}{a} \left( 1 + h_{00} + h_{0\alpha} h^{0\alpha} \right) \\
&\quad- (n-1) \frac{H}{a} \left( 2 \partial^\alpha h_{0\alpha} - \partial_0 h \right) - (n-1) \frac{H}{a} h^{0\alpha} \left( 2 \partial^\beta h_{\alpha\beta} - \partial_\alpha h \right) \\
&\quad+ (n-1) \frac{H}{a} h^{\alpha\beta} \left( 2 \partial_\alpha h_{0\beta} - \partial_0 h_{\alpha\beta} \right) \\
&\quad+ \frac{1}{4} a^{-2} \left[ 2 \left( \partial_\alpha h_{\mu\nu} \right) \partial^\alpha h^{\mu\nu} - 3 \left( \partial^\mu h_{\alpha\mu} \right) \partial_\nu h^{\alpha\nu} - \left( \partial_\alpha h_{\mu\nu} \right) \partial^\mu h^{\nu\alpha} + 2 \left( \partial_\alpha h \right) \partial_\nu h^{\alpha\nu} \right] \\
&\quad+ \frac{1}{4} a^{-2} \left[ \left( \partial^\alpha h - \partial^\mu h^\alpha_\mu \right) \left( \partial^\nu h_{\alpha\nu} - \partial_\alpha h \right) - \left( \partial^\alpha h^\mu_\nu - \partial^\mu h^\alpha_\nu \right) \left( \partial^\nu h_{\alpha\mu} - \partial_\alpha h^\nu_\mu \right) \right] \\
&\quad- 2 a^{-2} \partial_{[\mu} \partial^{[\mu} h^{\nu]}_{\nu]} - 2 a^{-2} h^{\alpha[\mu} \left( \partial^{\nu]} \partial_{[\mu} h_{\nu]\alpha} - \partial_\alpha \partial_{[\mu} h^{\nu]}_{\nu]} \right) + \bigo{h^3} \eqend{.}
\end{spliteq}
The $n$-dimensional Weyl tensor is defined by
\begin{equation}
\tilde{C}^{\mu\nu}{}_{\rho\sigma} \equiv \tilde{R}^{\mu\nu}{}_{\rho\sigma} - \frac{4}{n-2} \tilde{R}^{[\mu}_{[\rho} \delta^{\nu]}_{\sigma]} + \frac{2}{(n-1)(n-2)} \tilde{R} \delta^{[\mu}_{[\rho} \delta^{\nu]}_{\sigma]} \eqend{,}
\end{equation}
and to first order in perturbations we have
\begin{spliteq}
\tilde{C}^{\mu\nu}{}_{\rho\sigma} &= \frac{2}{(n-2) a^2} \bigg[ \partial_{[\rho} \partial^{[\mu} h - 2 \partial_\alpha \partial_{[\rho} h^{\alpha[\mu} + \partial^2 h^{[\mu}_{[\rho} + \frac{1}{n-1} \left( \partial_\alpha \partial_\beta h^{\alpha\beta} - \partial^2 h \right) \delta^{[\mu}_{[\rho} \bigg] \delta^{\nu]}_{\sigma]} \\
&\quad- 2 a^{-2} \partial_{[\rho} \partial^{[\mu} h^{\nu]}_{\sigma]} + \bigo{h^2} \eqend{.}
\end{spliteq}

\bibliography{literature}

\end{document}